\documentclass[aps, prx, 10pt, twocolumn,superscriptaddress]{revtex4-2}

\usepackage{graphicx}
\usepackage{color}
\usepackage{amsmath}
\usepackage{hyperref}

\newcommand{\mat}{NaGdP$_2$O$_7$}
\newcommand{\TN}{$T_{\rm N}$}
\newcommand{\mB}{$\mu_{\rm B}$}

\begin{document}
\title{Adibatic demagnetization refrigeration with antiferromagnetically ordered NaGdP$_2$O$_7$}

\author{P. Telang}
 \affiliation{EP VI, Center for Electronic Correlations and Magnetism, Institute of Physics, University of Augsburg, D-86159 Augsburg, Germany}

\author{T. Treu}
 \affiliation{EP VI, Center for Electronic Correlations and Magnetism, Institute of Physics, University of Augsburg, D-86159 Augsburg, Germany}
 
\author{M. Klinger}
 \affiliation{EP VI, Center for Electronic Correlations and Magnetism, Institute of Physics, University of Augsburg, D-86159 Augsburg, Germany}

\author{A. A. Tsirlin}
\affiliation{Felix Bloch Institute for Solid-State Physics, University of Leipzig, 04103 Leipzig, Germany}

\author{P. Gegenwart}
 \affiliation{EP VI, Center for Electronic Correlations and Magnetism, Institute of Physics, University of Augsburg, D-86159 Augsburg, Germany}

\author{A. Jesche}
 \email[]{anton.jesche@physik.uni-augsburg.de}
 \affiliation{EP VI, Center for Electronic Correlations and Magnetism, Institute of Physics, University of Augsburg, D-86159 Augsburg, Germany}

\begin{abstract}
We present a comprehensive study of the structural, magnetic, and thermodynamic properties, as well as the adiabatic demagnetization refrigeration (ADR) performance of \mat.
Although NaGdP$_2$O$_7$ exhibits antiferromagnetic ordering at a Néel temperature of $T_{\rm N} = 570$\,mK in zero field, ADR experiments achieved a minimum temperature of 220\,mK starting from $T = 2$\,K under an applied magnetic field of $\mu_0H = 5$\,T. 
The warm-up time back to $T = 2$\,K exceeds 60 hours with 9 hours below $T = 400$\,mK underscoring the potential of \mat~ as an efficient precooling stage in double-stage ADR systems. 
We show that NaGdP$_2$O$_7$ can be seen as a network of ferromagnetic spin chains with antiferromagnetic interchain couplings and also investigate the influence of antiferromagnetic ordering on the magnetic entropy. 
We find that the temperature dependence of the entropy plays a more dominant role than its magnetic field dependence in the magnetically ordered state. 
\end{abstract}

\maketitle

\section{Introduction}
Traditionally, adiabatic demagnetization of a paramagnetic salt is employed to obtain temperatures below $T = 0.5$\,K\,\cite{Giauque1933}. 
Later, stable and UHV-compatible compounds with large magnetic moments and extremely low ordering temperatures -- such as gadolinium gallium garnet Gd$_3$Ga$_5$O$_{12}$\,\cite{Schiffer1994} -- were introduced alongside the paramagnetic salts, which contained water molecules and were not UHV-compatible.
These compounds have a large fraction of nonmagnetic atoms to prevent magnetic ordering by exchange or dipole-dipole interactions. 
Currently the $^3$He/$^4$He dilution refrigeration technique is still the preferred method for achieving low temperatures due to its ability to continuously maintain sub-K temperatures and the absence of magnetic stray fields. 
However, the increasing cost of the rare and strategically important $^3$He isotope has spurred the search for new adiabatic demagnetization refrigeration (ADR) materials that would be capable of maintaining low temperatures for extended periods of time\,\cite{treu_utilizing_2025}. 
An ideal candidate for ADR would have high magnetic volume density, low magnetic anisotropy, and low magnetic ordering temperatures. 
A combination of these properties can be found in frustrated magnets, where the magnetic ions are more densely packed\,\cite{zhitomirsky_enhanced_2003}.
This idea has been taken up in various current research projects, see for example Refs.\,\onlinecite{Tokiwa2021, delacotte_nagds2_2022, xu_gdohf2_2022, koskelo_free-spin_2022, kleinhans_magnetocaloric_2023, yang_exceptional_2024}.   
Recently, remarkable ADR performance was observed in a spin supersolid candidate\,\cite{xiang_giant_2024}.

In our previous study, we investigated the ADR performance of NaYbP$_2$O$_7$, which achieved temperatures as low as $T_{\rm min} = 45$\,mK\,\cite{Arjun2023a}. 
However, the hold time -- defined as the duration between reaching $T_{\rm min}$ and returning to the initial temperature of $T = 2$\,K -- was limited to approximately one hour. This limitation arises from the low entropy of NaYbP$_2$O$_7$, constrained by the \mbox{pseudospin-1/2} nature of the Yb$^{3+}$ ion in the presence of a crystal electric field, yielding the maximum entropy of $R\ln2$ at low temperatures.

By replacing Yb$^{3+}$ with Gd$^{3+}$ ($S = 7/2$), the entropy can be increased to $R\ln8$, thus leading to enhanced magnetic volume density and potentially superior ADR performance. 
Here, we report the structure, thermodynamic behavior, and ADR performance of \mat. 
We show that this compound undergoes antiferromagnetic ordering at a Néel temperature of $T_{\rm N} = 570$\,mK, which is an order of magnitude higher compared to the Yb counterpart. 
Notwithstanding this fact, \mat~ demonstrates an impressive hold time of up to 63 hours making it an exceptional candidate for use as a precooler in two-stage ADR systems.

\section{Experiment}
%\subsection{Synthesis}
Polycrystalline \mat~was synthesized via a conventional solid-state reaction of Gd$_2$O$_3$, Na$_2$CO$_3$, and (NH$_4$)$_2$HPO$_4$. 
The precursors were thoroughly ground in an agate mortar and heated at 200$^\circ$C for 12\,h. 
The mixture was ground again and sintered in the temperature range of 500-650$^\circ$C. 
The temperature was increased in steps of 50$^\circ$C and each sintering step
lasted 12\,h. 
Increasing the temperature beyond 650$^\circ$C resulted in a decomposition of \mat~into GdPO$_4$ and NaPO$_3$.
Note that we have performed several similar attempts to grow iso-electronic KGdP$_2$O$_7$ in order to analyze the effect of replacing the alkaline metal similar to the investigation of $A$YbP$_2$O$_7$ ($A$ = Na, K)\,\cite{Arjun2023a}.
However, KGdP$_2$O$_7$ does not form for $T < 600^\circ$C and at higher temperatures GdPO$_4$ and KPO$_3$ formed instead.

%\subsection{X-Ray powder diffraction}
Phase purity of the sample was confirmed by X-ray powder diffraction (XRD) using a  PANalytical Empyrean diffractometer with Cu-K$_\alpha$ radiation at room temperature. 
Rietveld refinement of the observed XRD patterns was performed using the FullProf package\,\cite{Carvajal1993}.
The illustration of the crystal structure was generated using CrystalMaker.

The temperature-dependent heat capacity, $C(T)$, was measured using the heat capacity option of a Physical Property Measurement System (PPMS) Dynacool manufactured by Quantum Design equipped with a He3-option. 
%For the low temperature (0.4 - 2 K) Cp measurements the 3He option was used. 
In order to ensure strong thermal coupling, the specific heat measurements were performed on a pellet made by mixing an equivalent mass ratio of \mat~and silver powder. 
To extract the specific heat of the sample, the contribution from the silver powder was subtracted from the data based on measurements of pure silver pellets of comparable size.
Note that reported values for entropy and entropy density refer to \mat~only.
In order to estimate the phonon contribution, the $C(T)$ data at high temperatures ($14\,{\rm K} < T < 30$\,K) were fitted using a polynomial $C(T) = aT^3 + b T^5 + c T^7$ with $b$ and $c$ accounting for deviations from the low-temperature Debye approximation.
Magnetization as a function of temperature ($T = 0.4 - 300$\,K) and field (up to $\mu_0H = 7$\,T) was measured using a MPMS3 magnetometer manufactured by Quantum Design equipped with a He3-option. 
These measurements were performed on a silver-\mat~pellet similar to the one used for heat capacity measurements, with only a portion of the pellet utilized at low temperatures due to the limited range of the MPMS3.

For the ADR experiment, we used a 3.3\,g cylindrical pellet of 15\,mm diameter and 5.5\,mm thickness containing equal weights of \mat~and silver powder (average grain size of $\approx 1\,\mu$m). 
%Due to the insulating nature of NaGdP2O7, it is crucial to use silver powder for better thermal conductivity within the pellet. 
The pressed pellet was sintered at $600^\circ$C to further improve the thermal conductivity. 
The sample temperature was measured using a custom-built thermometer based on a commercial ruthenium oxide chip resistor. The substrate thickness was reduced by approximately 80\% to improve thermal coupling. This thermometer was calibrated against a known reference thermometer and read out with a Lake Shore Model 372 AC resistance bridge, equipped with a Model 3726 scanner, operating at a constant current of 1 nA. 
It was directly mounted onto the silver-\mat~pellet using GE varnish, with superconducting NbTi wires used for electrical contacts.
A (quasi-)adiabatic state is achieved by evacuating the sample chamber to a high vacuum of less than 1.3\,mPa once the initial state at $T = 2$\,K and $\mu_0H = 5$\,T is reached.

The ADR experiments were performed as reported previously\,\cite{Tokiwa2021, Arjun2023a, treu_utilizing_2025}. 
This facilitates a direct comparison of the ADR performances of this diphosphate with its Yb analogue.

Magnetic couplings $J_{ij}$ entering the spin Hamiltonian,
\begin{equation}
 \mathcal H=\sum_{\langle ij\rangle} J_{ij}\mathbf S_i\mathbf S_j
\end{equation}
where the summation is over the bonds $\langle ij\rangle$ and $S=7/2$, were obtained by the mapping analysis~\cite{xiang2011} using density-functional (DFT) band-structure calculations performed in the FPLO code~\cite{fplo} with the Perdew-Burke-Ernzerhof approximation~\cite{pbe96} for the exchange-correlation potential. Total energies of several spin configurations were converged down to $10^{-7}$\,Ha to ensure sufficient accuracy for the evaluation of weak exchange couplings between the Gd$^{3+}$ ions. Correlation effects in the Gd$^{3+}$ $4f$ shell were treated on the mean-field DFT+$U$ level with the on-site Coulomb repulsion $U=10$\,eV and Hund's coupling $J_H=1$\,eV. Temperature-dependent magnetic susceptibility and field-dependent magnetization of the resulting spin model were obtained from quantum Monte-Carlo (QMC) simulations using the \texttt{loop} algorithm~\cite{loop} of ALPS~\cite{alps}.

\section{Results}

\subsection{Crystal structure}

\begin{figure}
\includegraphics[width=0.46\textwidth]{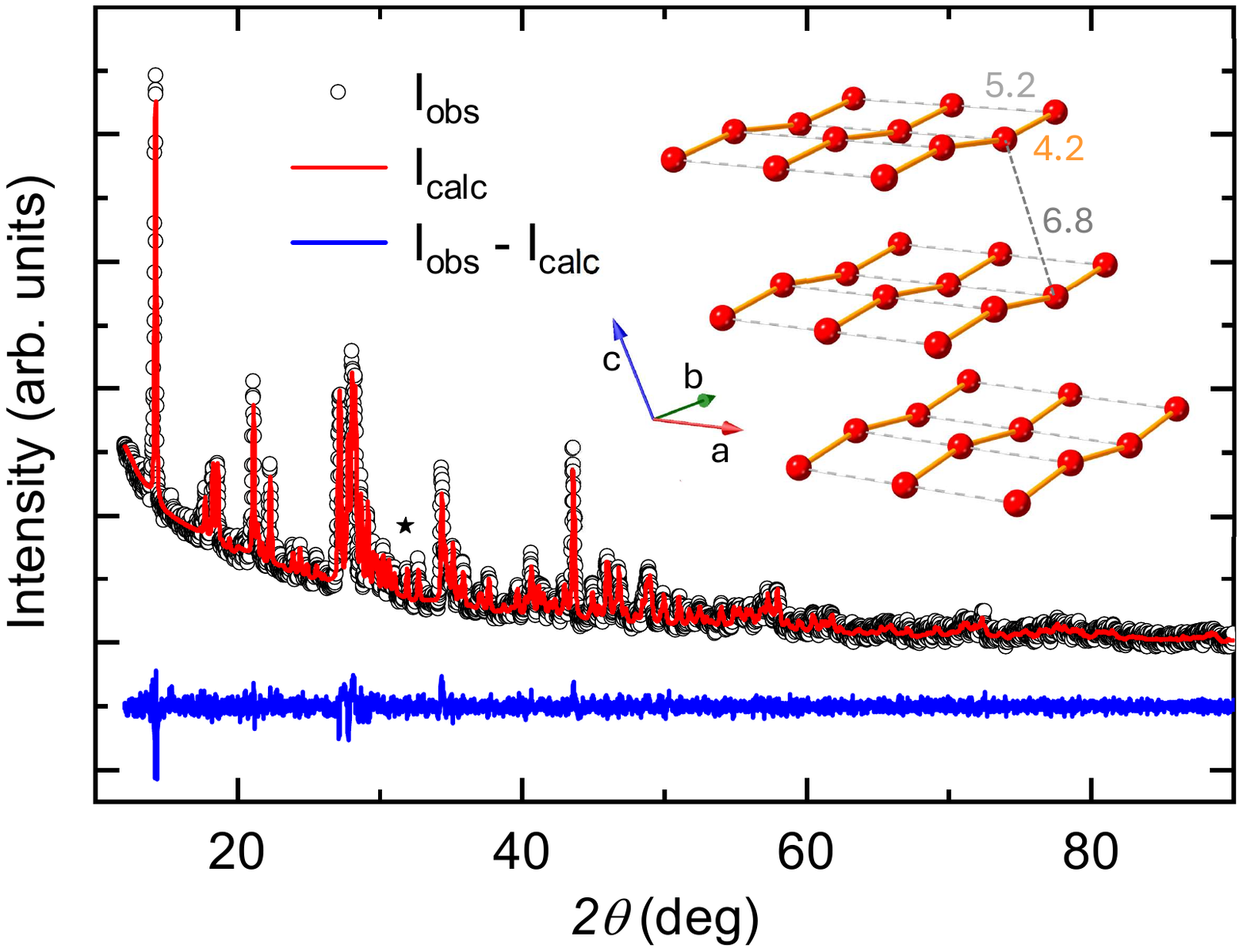}
\caption{X-Ray powder diffraction pattern of \mat~(Cu-$K_\alpha$). The refinement includes the minor impurity of GdPO$_4$ ($\approx 4.5$\,wt.\%). The largest peak of the impurity phase is marked by a star. 
The inset shows the arrangement of the Gd atoms with the distances given in units of \AA.
\label{xrd}}	
\end{figure}

\mat~crystallizes in a monoclinic lattice with the space group $P\,2_1/n$ (space group \#14)\,\cite{anisimova_double_1988} and is isostructural to NaEuP$_2$O$_7$\,\cite{Ferid2004}. 
The structure is built from GdO$_8$ polyhedra, slightly distorted PO$_4$ tetrahedra, and NaO$_8$ polyhedra. The GdO$_8$ polyhedra form infinite chains along the crystallographic $b$-axis, which are not connected along other directions\,\cite{zhu_crystal_2008}, suggesting the reduced dimensionality of the magnetic subsystem.
The positions of the Gd atoms are depicted in Fig.\,\ref{xrd} with the characteristic distances given in units of \AA.

\begin{figure}
\includegraphics[width=0.42\textwidth]{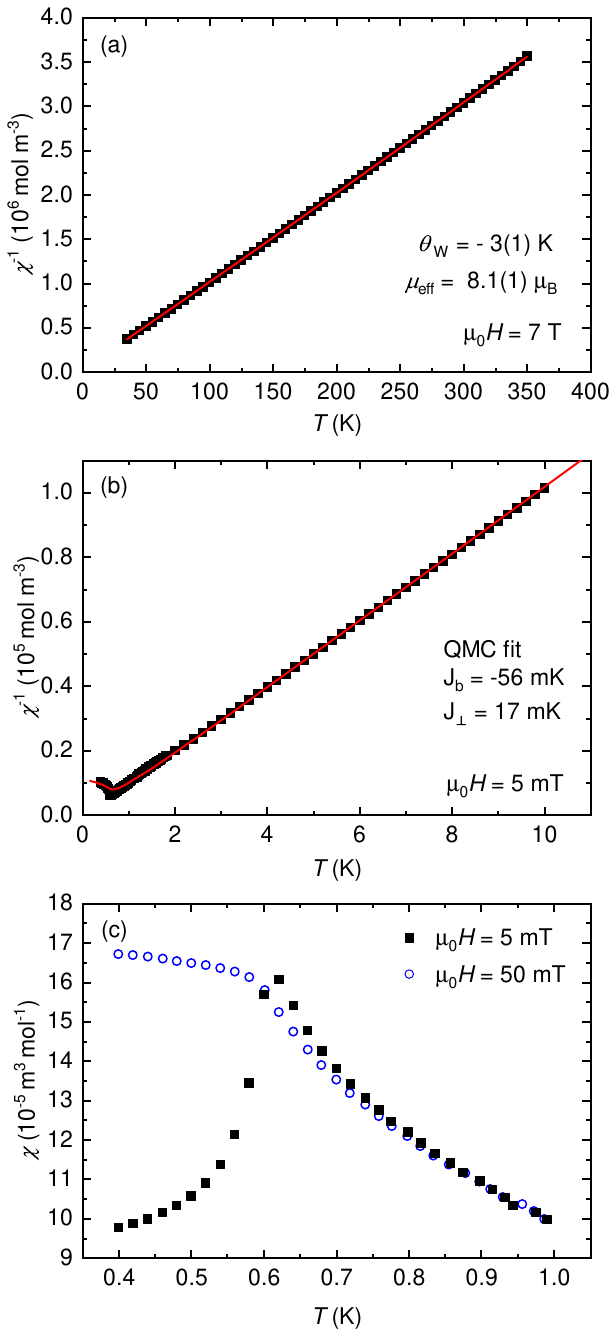}
\caption{Temperature-dependent magnetic susceptibility $\chi = M/H$ of \mat. 
(a) Curie-Weiss behavior in $1/\chi$. 
(b) Low-temperature susceptibility and its fit (red line) with the model of FM spin chains ($J_b=-56$\,mK) coupled by weaker AFM interchain couplings ($J_{\perp}=17$\,mK). The fitted curve is obtained by QMC and almost identical to a Curie-Weiss fit for $T > 1$\,K.
% is observed over the full temperature range $T = 1-350$\,K with an effective moment $\mu_{\rm eff} = 8.1\,\mu_{\rm B}$ and a small negative Weiss temperature $\Theta_{\rm W}$. 
(c) AFM ordering is indicated by the sharp peak centered at $T = 0.6$ \,K.
 \label{chiT}}	
\end{figure}

The room-temperature XRD powder pattern of \mat~is shown in Fig.\,\ref{xrd}, along with the Rietveld refinement.
The measured diffraction pattern is well described based on the reported unit cell parameters\,\cite{anisimova_double_1988}.
Note that there is a contradicting report that assigns a different, orthorhombic lattice\,\cite{Kizilyalli1993}.
As the pattern showed additional peaks corresponding to the GdPO$_4$ impurity, a two-phase refinement was performed with a goodness of fit of $\chi^2 = 1.8$. 
We have found a concentration of 4.5\,wt.\% GdPO$_4$ (which orders antiferromagnetically below $T = 0.8$\,K\,\cite{palacios_magnetic_2014}).
The obtained lattice parameters of \mat~are 
$a = 5.222$\,\AA, 
$b = 8.419$\,\AA, 
$c = 13.392$\,\AA, 
and $\beta = 111.6$.
Those values are in a good agreement with the literature data\,\cite{anisimova_double_1988}.

\subsection{Magnetic susceptibility}
\label{sec:chi}

The high-temperature part of the magnetic susceptibility, $\chi=M/H$, is shown in Fig.\,\ref{chiT}a ($T = 30-350$\,K). 
A Curie-Weiss behavior is observed over the full temperature range. 
A fit to $\chi(T) = C/(T-\Theta_{\rm W})+\chi_0$ yields an effective moment of $\mu_{\rm eff} = 8.1\,\mu_{\rm B}$, a Weiss temperature of $\Theta_{\rm W} = -3$\,K, and a temperature-independent offset of $\chi_0 = -7.5 \times 10^{-9}$m$^3$mol$^{-1}$ (2\% of the room-temperature susceptibility, primarily caused by Larmor diamagnetism).
The effective moment is slightly increased compared to the value of the free Gd$^{3+}$ ion ($\mu_{\rm eff} = 7.94\,\mu_{\rm B}$), which is probably caused by the neglect of the sample geometry. 

The low-temperature behavior of $\chi(T)$ (Fig.\,\ref{chiT}b) shows that the linear evolution of $1/\chi$ persists down to 3\,K with the Curie-Weiss fit returning $\Theta_{\rm W} = -0.1(3)$\,K, and only at lower temperatures a weak curvature appears, followed by the sharp peak in $\chi(T)$ centered at $T = 0.6$\,K (Fig.\,\ref{chiT}c). 
The sharp decrease of $\chi(T)$ below this temperature as well as the strong suppression of the transition in somewhat larger applied fields indicates an antiferromagnetic (AFM) ordering. 
The AFM ordering temperature is estimated to $T = 580(20)$\,mK based on the maximum in the Fisher's heat capacity, d$(\chi T)$/d$T$\,\cite{Fisher1962}. 

In order to analyze the origin of this transition, we calculated exchange coupling in \mat~ using DFT. We find the dominant ferromagnetic (FM) coupling $J_b=-70$\,mK along the structural chains of the GdO$_8$ polyhedra. 
The couplings along the $a$-direction are much weaker and AFM, $J_a=5$\,mK. Two nonequivalent exchange pathways exist for the Gd--Gd interactions along $c$ (6.76\,\r A and 6.86\,\r A, respectively) because of the shift of the adjacent chains relative to each other. The averaged exchange coupling along these pathways, $\bar J_c=5$\,mK, is also AFM and comparable to $J_a$ in magnitude. 
The above exchange couplings are augmented by dipolar couplings between the Gd$^{3+}$ ions. 
From the interatomic distances we estimate $|J_b^D|=26$\,mK, $|J_a^D|=14$\,mK, and $|J_c^D|=6$\,mK obtained as the dipolar coupling energies divided by $S(S+1)$.
Therefore, the spin lattice of \mat~ should comprise FM spin chains with predominantly (super)exchange couplings. 
These chains are coupled antiferromagnetically, with the interchain couplings being largely dipolar in nature.

The magnetic behavior of \mat~ can be described by the model of FM spin chains ($J_b<0$) with weaker AFM interchain couplings.
For simplicity we choose the single value of $J_\perp$ that corresponds to $J_a+J_a^D$ (two  interchain couplings along $a$ per site) and $2 (J_c+J_c^D)$ (four interchain couplings along $c$ per site, see Fig.\,\ref{xrd}, from one layer to the next one the chains are shifted along $a$ by roughly $a/2$ with respect to $c^*$). 
This simplified model allows a reasonable description of the magnetic susceptibility [see Fig.~\ref{chiT}(b)] using $J_b=-56$\,mK and $J_{\perp}=17$\,mK ($g=1.97$) that also reproduce the AFM ordering temperature of about 600\,mK. 
The fitted $J$-values are in a quite good agreement with the DFT results because $J_{\perp}$ should be seen as average of $J_a+J_a^D$ and $2(J_c+J_c^D)$.

\subsection{Isothermal magnetization}

\begin{figure}
\includegraphics[width=0.46\textwidth]{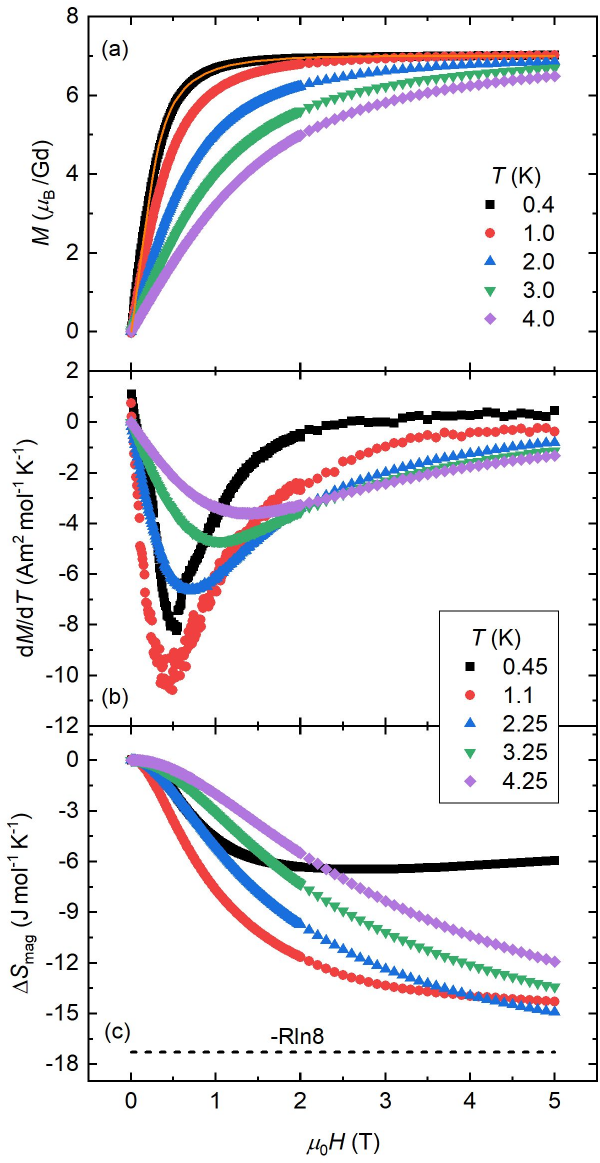}
\caption{a) Isothermal magnetization of \mat~expressed in Bohr magneton per Gd$^{3+}$. The orange line shows the simulated magnetization curve at $T = 0.4$\,K obtained by QMC for the effective model introduced in Sec.~\ref{sec:chi}. 
b) Derivative of the magnetization with respect to temperature as a function of the applied magnetic field.
c) Change of the magnetic entropy with applied field calculated by integrating d$M$/d$T$ over $\mu_0H$.
 \label{mh}}	
\end{figure}

The isothermal magnetization $M(H)$ is shown in Fig.\,\ref{mh}a. 
For the largest applied fields of $\mu_0H = 5$\,T, the measured moments at $T < 2$\,K approach their saturation value and are in a good agreement with the saturation moment of the free ion ($\mu_{\rm sat} = 7.0\,\mu_B$). 
The model of FM spin chains with weaker AFM interchain couplings correctly describes the field evolution of the magnetization at 0.4\,K.

Fig.\,\ref{mh}b shows the change of $M$ with respect to temperature, d$M$/d$T$, as a function of field. 
The derivatives were approximated by subtracting $M(H)$ curves obtained at adjacent temperatures (separated by $\Delta T$) and dividing by $\Delta T$. 
The following pairs of $M(H)$ curves were used; 
0.4\,K and 0.5\,K $\rightarrow$ d$M$/d$T$ at $T = 0.45$\,K,  
1.0\,K and 1.2\,K $\rightarrow$ d$M$/d$T$ at $T = 1.1$\,K, 
2.0\,K and 2.5\,K $\rightarrow$ d$M$/d$T$ at $T = 2.25$\,K,
and similarly for d$M$/d$T$ at $T = 3.25$\,K and 4.25\,K.

The change in d$M$/d$T$ is largest at intermediate temperatures of $T \approx 1$\,K due to the rapid saturation at lower temperatures and the smaller magnetization at higher temperatures.  
Experimental d$M$/d$T$ allows one to calculate the difference in the magnetic entropy by taking advantage of the basic Maxwell relations that lead to
\begin{equation}
\Delta S_{\rm mag}(T,H) = \mu_0 \int_0^H \frac{dM(T,H)}{dT} dH
\end{equation}

The obtained values are plotted in Fig.\,\ref{mh}c.
As expected, the entropy decreases with increasing applied field, which is shown by the negative sign of $\Delta S_{\rm mag}(T,H)$.
At the lowest temperature of $T = 0.45$\,K, we find $\Delta S_{\rm mag}(T,H) = -6.4$\,J\,mol$^{-1}$\,K$^{-1}$ for $\mu_0H = 2$\,T. 
The slight increase of $\Delta S_{\rm mag}(T,H)$ with further increasing field is an artifact caused by the measurement error of $M(H)$.
Larger changes in entropy are observed at $T = 1.1$\,K with $\Delta S_{\rm mag}(T,H) = -14.2$\,J\,mol$^{-1}$\,K$^{-1}$
For higher temperatures, $\Delta S_{\rm mag}(T,H)$ keeps increasing with increasing field even for the largest applied fields and asymptotically approaches the maximum possible value of $R\ln8$ = 17.3\,J\,mol$^{-1}$\,K$^{-1}$.
The smaller changes in the entropy at the lowest temperature are caused by the lower zero-field entropy (see below), which limits the maximum possible entropy change that can be achieved by increasing $H$. 

\subsection{Heat capacity}

\begin{figure}
\includegraphics[width=0.46\textwidth]{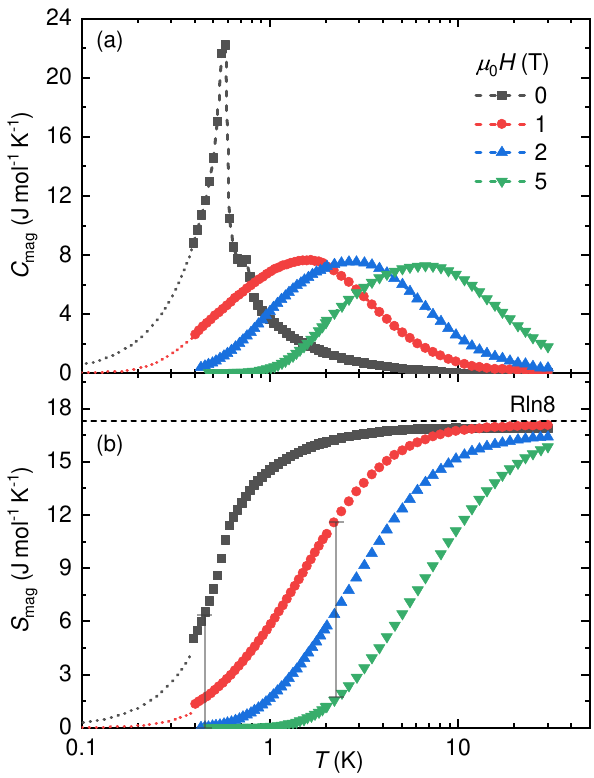}
\caption{a) Temperature-dependent magnetic heat capacity $C_{\rm mag}(T)$ of \mat. The dotted lines denotes extrapolated values for $T \rightarrow 0$, the dashed line is a guide to the eye. 
The sharp lambda-type anomaly in $H = 0$ indicates AFM ordering at $T_{\rm N} = 570$\,mK.
b) Magnetic entropy $S_{\rm mag}$ calculated from $C(T)$. 
For $\mu_0H = 0$\, and 1\,T, the entropy for $T < 0.4$\,K was calculated based on $\Delta S_{\rm mag}$ obtained from field-integration of $dM/dT$ (vertical lines).
The dotted lines denote magnetic entropies estimated from extrapolated $C_{\rm mag}(T)$ data.
 \label{hc}}	
\end{figure}

The temperature-dependent, magnetic contribution to the heat capacity, $C_{\rm mag}(T)$, is shown in Fig.\,\ref{hc}a.
A sharp, lambda-type anomaly for $H = 0$ marks the AFM ordering at $T_{\rm N} = 570(12)$\,mK in good agreement with $\chi(T)$.
The black, dotted line marks an extrapolation for $T \rightarrow 0$ based on an (assumed) power-law dependence $C \sim T^2$ for $T < 0.5$\,K. 
Another, small anomaly at $T = 0.74$\,K is attributed to the AFM ordering of the impurity phase GdPO$_4$ in accordance with the previous results\,\cite{Thiriet2005}.
The entropy associated with this transition amounts to 3.5\% of the total magnetic entropy in a reasonable agreement with the phase fraction estimated by XRD.

Further measurements of $C_{\rm mag}(T)$ were performed in applied fields of $\mu_0H = 1, 2,$ and 5\,T.
In $\mu_0H = 1$\,T, the AFM ordering is strongly suppressed and a Schottky-type anomaly appears. 
An extrapolation of $C_{\rm mag}(T)$ similar to the one for $H = 0$ was performed for the data obtained in $\mu_0H = 1$\,T with an additional Schottky-type contribution (red, dotted line).
At the lowest measured temperature of $T = 0.4$\,K, the heat capacity is negligible for $\mu_0H = 2$\,T and 5\,T.
With increasing applied field, the maximum of the Schottky-type anomalies is shifted to higher temperatures.

Fig.\,\ref{hc}b shows the magnetic entropy calculated from $C_{\rm mag}(T)$. 
Below $T = 0.4$\,K, the lowest measured temperature, there is a significant amount of entropy for $H = 0$ that is not accessible from $C_{\rm mag}(T)$.  
Therefore, we took advantage of the entropy difference calculated from the above magnetization analysis. 
Since $S_{\rm mag}$ is essentially zero for $T = 0.45$\,K in $\mu_0H = 5$\,T, the calculation of $\Delta S_{\rm mag} = S_{\rm mag}$(0T)$ - S_{\rm mag}$(5T)$ = 6.4$\,J\,mol$^{-1}$\,K$^{-1}$ even yields the absolute value of $S_{\rm mag}$ for $H = 0$.
Note that $S_{\rm mag}$ is strongly temperature-dependent in this temperature range and increases by 1.38\,J\,mol$^{-1}$\,K$^{-1}$ between $T = 0.40$\,K and $T = 0.45$\,K.
In this respect, the agreement between the obtained total entropy of $S_{\rm mag} = 16.9$\,J\,mol$^{-1}$\,K$^{-1}$ (at $T = 30$\,K) and the expected value of $R\ln8 = 17.3$\,J\,mol$^{-1}$\,K$^{-1}$  can be regarded as exceptionally good. 
The black, dotted line between $T = 0.1$\,K and 0.4\,K denotes the low-temperature entropy estimated from the extrapolated $C_{\rm mag}(T)$ data. 
The obtained values are somewhat smaller than the ones based on the $M(H)$ data. 

Entropy is shifted towards higher temperatures in the presence of applied magnetic fields.
For $\mu_0H = 1$\,T, the low-temperature contribution to $S_{\rm mag}$ was calculated from $M(H)$ similar to the case of $H = 0$.
Since $\Delta S_{\rm mag} = S_{\rm mag}{\rm (1T)} - S_{\rm mag}{\rm(5T)} = 9.85$\,J\,mol$^{-1}$\,K$^{-1}$ at $T = 2.25$\,K is larger than the corresponding entropy difference at $T = 0.45$\,K, the estimation was done based on the values at the higher temperature (though the calculation leads to similar results when using $\Delta S_{\rm mag}$ for lower temperatures). 
The relative difference to $S_{\rm mag}$ calculated from the extrapolated $C(T)$ data (red, dotted line) is somewhat larger than for $H = 0$ and indicates a possible underestimation of the low temperature specific heat caused by the competition between AFM exchange and Zeeman energy.
The magnetic entropy amounts to $S_{\rm mag} = 17.1$\,J\,mol$^{-1}$\,K$^{-1}$ (at $T = 30$\,K) in excellent agreement with $R$ln8.
This corresponds to a volumetric entropy density of 210\,mJ\,K$^{-1}$cm$^{-3}$.
For larger applied fields, $S_{\rm mag}$ approaches $R\ln8$ and at $T = 30$\,K reaches 95\% and 91\% of $R\ln8$ for $\mu_0H = 2$\,T and 5\,T, respectively.

\subsection{Adiabatic demagnetization refrigeration}
\begin{figure}
\includegraphics[width=0.45\textwidth]{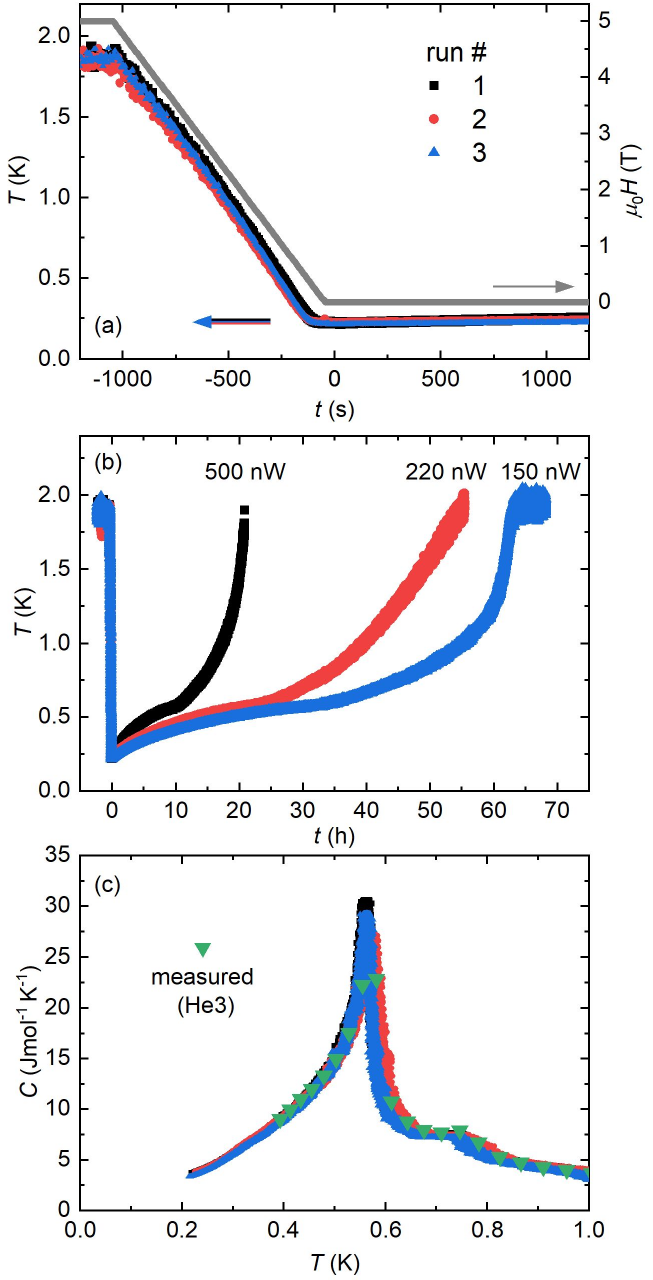}
\caption{Adiabatic demagnetization refrigeration of \mat~starting from $T = 2$\,K and $\mu_0H = 5$\,T. A pressed pellet of 50 wt.\% Ag and 50 wt.\% \mat~was measured three times.
a) minimum temperature of $T_{\rm min} = 220$\,mK is achieved after ramping the field to $H = 0$.
b) The warming curve strongly depends on the vacuum level in the sample chamber that determines the heat input through convection by the residual gas. The numbers give the calculated heat input per time $\dot{Q}$ obtained from a comparison of the specific heat data, which is shown in c): heat capacity calculated from ADR by $\dot{Q} = C_{\rm ADR}\dot{T}$ and directly measured in the PPMS.
 \label{adr}}	
\end{figure}

The main results of the ADR performance for \mat~are displayed in Fig.\,\ref{adr}. 
The starting temperature was $T = 2$\,K (slightly shifted due to the neglect of the magnetoresistance of the thermometer) with an initial applied field of $\mu_0H = 5$\,T. 
Upon ramping the field down to zero at a sweep rate of 5\,mT/s, the minimum temperature of $T_{\rm min} = 220(7)$\,mK was achieved consistently across three independent runs (see the time-dependent temperature $T(t)$ in Fig.\,\ref{adr}a, where $t = 0$ corresponds to reaching $T_{\rm min}$). 
The exact determination of $t = 0$ is challenging due to the minimal change in $T(t)$ around this point. 
The data suggests a delay of approximately 45 seconds between reaching $H = 0$ and the lowest temperature.
Following this, the temperature increased linearly by 27 mK in run 1 and 12 mK in runs 2 and 3 over the first 1000\,s. 
All three runs were conducted under nominally identical conditions, using the same sample and equipment. 
Notably, the pressure sensor of the Dynacool PPMS indicated values below its minimum sensitivity of 1.3\,mPa.

Nevertheless, it became evident that the dominant factor in the heat input was most likely the presence of residual gas at the bottom of the sample chamber.
This conclusion is supported by the observed strong dependence of the warming curves on the interval between regenerating the cryopump and performing the ADR experiments, as shown in Fig.\,\ref{adr}b. 
Run 1, conducted six months after cryopump regeneration, exhibited markedly different warming behavior compared to run 2, which took place four weeks post-regeneration, and run 3, which occurred immediately after regeneration. 
The hold time was 21\,h, 56\,h, and 63\,h for runs 1, 2, and 3, respectively.
The numbers provided in this panel refer to the heat input per time, $\dot{Q}$ (implied from the heat capacity data, see Fig.\,\ref{adr}c and next paragraph) and run from $\dot{Q} = 150$ to 500\,nW.
Note that the heat input during a similar ADR run on the Yb-based analogue\,\cite{Arjun2023a} was roughly 700\,nW.  

Assuming $\Delta Q = C_{\rm ADR}\Delta T$ and accordingly $\dot{Q} = C_{\rm ADR}\dot{T}$, the heat capacity can be calculated from the time-derivative of $T(t)$ during warming. 
$\dot{Q}$ was chosen such that $C_{\rm ADR}(T)$ matches the values directly obtained by using the heat capacity option of the PPMS (see previous subsection). 
The results are shown in Fig.\,\ref{adr}c.
The sharp lambda-type anomaly at \TN, along with the subtle feature attributed to the GdPO$_4$ impurity, is evident in all three runs.
As indicated in Ref.\,\onlinecite{jesche_adiabatic_2023}, it seems surprising to achieve such an agreement by assuming a temperature-independent heat input $\dot{Q}$.

\subsection{Magnetic entropy in the vicinity of AFM ordering}

\begin{figure}
\includegraphics[width=0.47\textwidth]{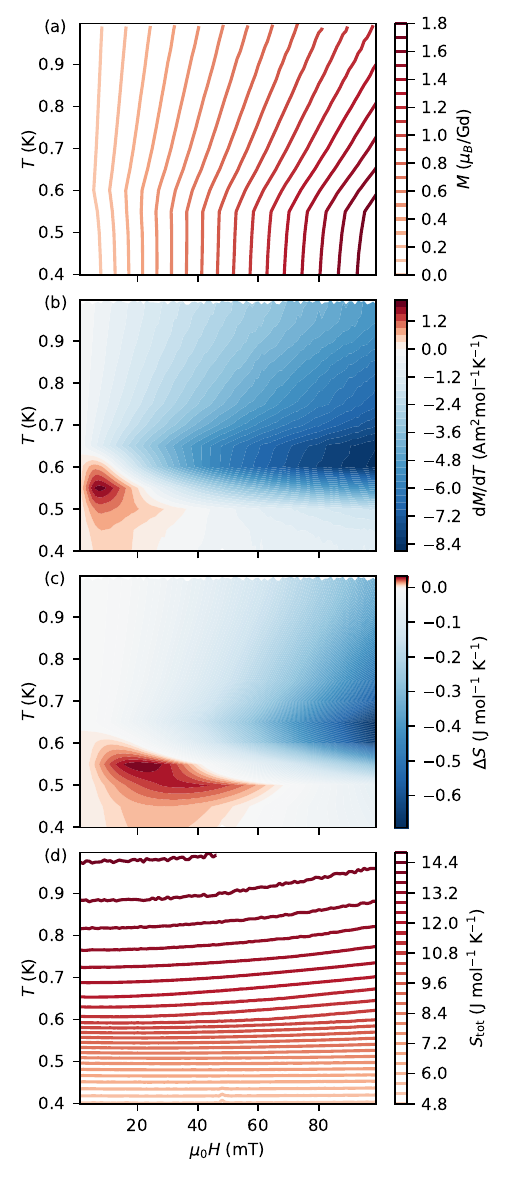}
\caption{Contours of the magnetization and derived quantities of \mat~in the vicinity of the AFM ordering as a function of temperature and applied field (same scale for all panels).
a) The AFM ordering in small applied fields is marked by a kink in the contour. 
b) Derivative of the magnetization with respect to temperature. In small fields at low temperature the magnetization increases with temperature due to the suppression of AFM order.
c) Change in entropy obtained by integrating d$M$/d$T$ over $\mu_0H$. The maximum in $\Delta S$ indicates the state of the largest, magnetic field-induced disorder.
d) Sum of temperature-dependent entropy obtained from $C(T, H = 0)$ and $\Delta S$.
 \label{adrafm}}	
\end{figure}

The AFM ordering at \TN~significantly affects the temperature and field dependence of the magnetic entropy and other thermodynamic quantities. 
The implications are difficult to trace in ADR experiments presented in the previous subsection due to the rapid changes in temperature during the field sweep.
Therefore, we employ careful magnetization measurements performed below $T = 1$\,K and analyze the entropy landscape as a function of field and temperature. 
To this end, the isothermal magnetization was measured from $T = 0.4$\,K to 1.0\,K in steps of $\Delta T = 0.05$\,K for fields from $\mu_0H = 0$ to 100\,mT in steps of 1\,mT.
Figure\,\ref{adrafm}a presents the magnetization data as a contour plot.
The largest obtained values of $\approx 1.8$\,\mB~per Gd in $\mu_0H = 100$\,mT amount to $\sim 25$\% of the saturation moment.
For non-interacting Gd moments, however, the expectation value is $6.3\,\mu_{\rm B}$ or 90\% of the saturation moment. 
The large difference between those values shows the strong effect of AFM interchain interactions and indicates that the critical field for suppressing the AFM order is significantly larger than $\mu_0H = 0.1$\,T. 
Indeed, using $J_{\perp}=17$\,mK from the susceptibility fit and $S=7/2$, one expects the saturation field of $\mu_0H_s = 2 z J_{\perp}S(k_B/g\mu_B)=355$\,mT ($z = 4$ nearest neighbors).

For small fields the AFM ordering is apparent from a clear kink in the contour, and the magnetization is increasing with temperature for low temperatures.
This is better seen in the derivative d$M$/d$T$ depicted in Fig.\,\ref{adrafm}b: positive values are observed in the vicinity of \TN.
There is a well-defined region around $T = 550$\,mK and $\mu_0 = 8$\,mT where the magnetization shows the strongest temperature dependence 
(note that this field is well below the remnant field of the magnet, which is smaller than 1\,mT for the applied fields used in those scans).
In intermediate fields, a sharp change in sign is seen above the critical temperature.
Nevertheless, this sign change does not mark the AFM phase boundary.
The small measured magnetization (see above) and the remaining anomaly at \TN~in $\mu_0H = 100$\,mT (Fig.\,\ref{adrafm}a) demonstrate the presence of AFM order in this field range.

The corresponding change in entropy with field, $\Delta S$, is obtained by an integration along the field axis (see also Fig.\ref{mh}c and text) with the result shown in Fig.\,\ref{adrafm}c.
Two well-separated regions are observed. The one at lower temperatures shows an increase of $S$ with $H$, whereas the one at higher temperatures shows the decrease.
The maximum entropy is accumulated in the region around $\mu_0H = 22$\,mT and $T = 550$\,mK and indicates the state of the largest, magnetic field-induced disorder.
The largest increase in the entropy $\Delta S$ (as a result of changing $H$) amounts to 32\,mJ\,mol$^{-1}$\,K$^{-1}$, which is more than two orders of magnitude smaller than the change in $S$ caused by increasing the temperature to above \TN~in $H = 0$.

The sum of $S(T,H = 0)$ obtained from $C(T)$ and $\Delta S(T,H)$ yields the total magnetic entropy, $S_{\rm tot}$, and is depicted in Fig.\,\ref{adrafm}d. 
It can be seen that the entropy landscape is dominated by the temperature dependence.
As expected, $S_{\rm tot}$ decreases with increasing field for $T > T_{\rm N}$.
For $T < T_{\rm N}$, the effect of $H$ is significantly smaller and the contribution of $\Delta S$ causes only a minor bending of the almost horizontal isentropes that is not recognizable in the depiction chosen. 

\section{Discussion}
First, we compare \mat~with several other potential ADR materials, such as NaYbP$_2$O$_7$\,\cite{Arjun2023a}, KBaGd(BO$_3$)$_2$\,\cite{jesche_adiabatic_2023}, and KBaYb(BO$_3$)$_2$\,\cite{Tokiwa2021} (see Tab.\,\ref{comp}). 
%that were investigated for their low temperature magnetic properties and ADR performance.
In the Yb-based compounds, no magnetic ordering was observed and ADR temperatures well below 50\,mK were found.
In contrast, AFM ordering occurs in the Gd-based compounds.
The N\'eel temperature, $T_{\rm N} = 570$\,mK, of \mat~is approximately twice that of KBaGd(BO$_3$)$_2$, which has $T_{\rm N} = 262$\,mK.

\begin{table}
\caption{\label{comp}
Comparison of ADR key parameters; ordering temperature $T_{\rm N}$, minimum ADR temperature achieved in PPMS setup $T_{\rm min}$, ground state entropy normalized by volume $S_{\rm GS}$.
}
\begin{tabular}{lccc}
\hline
\hline
~					&~~~$T_{\rm N}$~~~ 	&	$T_{\rm min}$ 	& $S_{\rm GS}$   \\
~					&~~~ (mK)~~~ 				& (mK)						& (mJ\,K$^{-1}$cm$^{-3}$)\\
\hline
KBaYb(BO$_3$)$_2$\,\cite{Tokiwa2021}			&	9$^*$	&	40	& 64		\\
KBaGd(BO$_3$)$_2$\,\cite{jesche_adiabatic_2023}	&	263		&	122	& 192		\\
NaYbP$_2$O$_7$\,\cite{Arjun2023a}				&	28$^*$	&	45	& 64		\\
NaGdP$_2$O$_7$ [this work]						&	570		&	220	& 210		\\
NaYbGeO$_4$\,\cite{Arjun2023b}					&	210		&	135	& 101		\\
\hline
\hline
\end{tabular}
\begin{flushleft}
$^*$ Estimated based on $T_{\rm N}$ of the Gd analogue,\\ see text and\,\cite{jesche_adiabatic_2023}.
\end{flushleft}
\end{table}

Surprisingly, \mat~exhibits a smaller (absolute) Weiss temperature of $\Theta_{\rm W} = -0.1(3)$\,K compared to KBaGd(BO$_3$)$_2$ with $\Theta_{\rm W} = -0.55(3)$\,K\,\cite{jesche_adiabatic_2023}, despite the opposite order of their Néel temperatures. Furthermore, the Weiss temperature $\Theta_{\rm W}$ of \mat~ is smaller than its Néel temperature $T_{\rm N}$, resulting in the frustration ratio of $\Theta_{\rm W}/T_{\rm N} < 1$. 
The reason for this counter-intuitive behavior is the interplay of FM intrachain and AFM interchain couplings in \mat\ that enter the Weiss temperature as a linear combination and bring $\Theta_{\rm W}$ close to zero.

Notably, we observe similar ratios of $r_{\rm ADR-AFM} = r_{\rm AA} = T_{\rm min}/T_{\rm N}$, which are 0.46 for KBaGd(BO$_3$)$_2$ and 0.39 for \mat. 
This ratio, $r_{\rm AA}$, can serve as an indicator of the performance of magnetically ordered ADR materials, as \TN $\cdot S$ offers a rough estimate of the energy needed to heat the system above \TN~and, therefore, significantly impacts the hold time (of course, an accurate calculation requires the precise distribution of $S$ over $T$).
It also explains the very different hold times between those two Gd-based materials despite their similar entropy densities (210\,mJ\,K$^{-1}$cm$^{-3}$ for \mat~and 192\,mJ\,K$^{-1}$cm$^{-3}$ for KBaGd(BO$_3$)$_2$\,\cite{treu_utilizing_2025}). 
In order to illustrate this, we scale the hold times to the same heat input rate of $\dot{Q} = 500$\,nW and obtain 8\,h\,$\cdot 710$\,nW/500\,nW\,$\approx 11$\,h in KBaGd(BO$_3$)$_2$~\cite{jesche_adiabatic_2023} vs. 21\,h in run 1 of \mat. The ratio of the resulting hold times (11\,h/21\,h) compares well to the ratio of the \TN's (263\,mK/570\,mK) despite the very different shapes of the AFM anomaly in the heat capacity.

The 210\,mJ\,K$^{-1}$cm$^{-3}$ entropy density of \mat~is comparatively large. 
Even higher values are found for GGG, GdPO$_4$, and Gd$_{9.33}$[SiO$_4$]$_6$O$_2$ that amount to 363\,\cite{kleinhans_magnetocaloric_2023}, 401\,\cite{palacios_magnetic_2014}, and 509\,\cite{yang_exceptional_2024}, respectively (in mJ\,K$^{-1}$cm$^{-3}$). 
The suitability for applications, however, is going to depend on the refrigerant capacity\,\cite{treu_utilizing_2025}, which is strongly temperature dependent, and various other factors such as thermal conductivity, thermal coupling, possible degassing, thermal cycling stability, etc. 

Similar to the case of KBaGd(BO$_3$)$_2$ and KBaYb(BO$_3$)$_2$, we estimate the ordering temperature of NaYbP$_2$O$_7$ based on the assumption that dipolar and exchange interactions scale with the square of the magnetic moment\,\cite{jesche_adiabatic_2023}.
Using the ratio $(\mu_{\rm sat}^{\rm Gd}/\mu_{\rm sat}^{\rm Yb})^2 = (7.0/1.55)^2 = 20$ and assuming comparable effective exchange paths, we estimate $T_{\rm N} = 28$\,mK.

We now turn to the impact of AFM ordering on the ADR process. 
While temperature decreases with {\it decreasing} applied magnetic field in a regular ADR process, a decrease in temperature is  expected when {\it increasing} the field from $H = 0$ at temperatures below $T_{\rm N}$ (see, for example, Ref.\,\cite{diederix_theoretical_1979}). This behavior has been partially discussed in the context of quantum criticality, where a corresponding decrease in temperature upon crossing the AFM phase boundary has been observed both experimentally~\cite{wolf_magnetocaloric_2011, gegenwart_gruneisen_2016} and theoretically\,\cite{garst_sign_2005, wolf_cooling_2014}.
ADR and the magnetocaloric effect in general in the presence of magnetic phase transitions is discussed for example in Refs.\,\onlinecite{tishin_magnetocaloric_2003, de_oliveira_theoretical_2010, gomez_magnetocaloric_2013} and references therein.
 
More recent results on low-temperature AFM ordering in connection to ADR are presented in Refs.\, \onlinecite{Xiang2023-arxiv, Arjun2023b, khatua_magnetic_2024, zeng_k2renb5o15_2024, xu_achieving_2024, treu_utilizing_2025}.
In particular, the calculation of $C_{\rm mag}(T)$ from ADR $T(t)$ curves was shown in comparison with directly measured heat capacity\,\cite{Arjun2023b, treu_utilizing_2025}. 
In line with general thermodynamic relations, the total entropy of \mat~does increase with increasing temperature despite the opposite contribution of the magnetic field induced $\Delta S$ as shown in Fig.\,\ref{adrafm}c,d.
This is fully consistent with the almost constant temperature observed in the ADR experiment for $\mu_0H < 0.3$\,T (Fig.\,\ref{adr}a).
On the other hand, it presents a significant difference to KBaGd(BO$_3$)$_2$, which showed a sizable increase in temperature during ADR for $\mu_0H < 0.3$\,T\,\cite{jesche_adiabatic_2023}.
Assuming that both ADR experiments follow roughly the same isentrope when starting at $T = 2$\,K and $\mu_0H = 5$\,T (when neglecting different magnetic exchange and heat capacity) and given that the magnetic phase boundary is crossed at roughly half of the respective \TN, it seems surprising to find such a difference in the AFM phase.
Considering the differences in the heat capacities of \mat~compared to KBaGd(BO$_3$)$_2$ does not provide an explanation, since at $T_{\rm min}$ \mat~possesses only 12\% of $C_{\rm mag}(T_{\rm N})$, whereas in KBaGd(BO$_3$)$_2$ this values increases to 24\% due to the broader distribution of $C_{\rm mag}(T)$.
Accordingly, a larger temperature dependence would be expected for \mat~when similar changes of the magnetic state variables are performed, at odds with the observed behavior.   
More detailed thermodynamic measurements of KBaGd(BO$_3$)$_2$ and related (Gd-based) materials (for example\,\cite{chen_growth_2024, zeng_k2renb5o15_2024}) in moderate applied fields are needed in order to achieve a better understanding of the low-temperature ADR in the vicinity of AFM ordering. 
It is also worth noting that \mat\ is a nonfrustrated quasi-one-dimensional magnet with leading FM couplings, whereas KBaGd(BO$_3$) is a quasi-two-dimensional frustrated antiferromagnet with possible effects of randomness.

\section{Summary}
\mat~shows a vanishingly low Weiss temperature of $\Theta_{\rm W} = -0.1(3)$\,K and orders antiferromagnetically at \TN~= 570\,mK.
A minimum temperature of $T_{\rm min} = 220$\,mK was reached in quasi-adiabatic demagnetization experiments with the hold time of up to 63\,h below $T = 2$\,K in a standard Dynacool PPMS (for initial $T = 2$\,K and $\mu_0H = 5$\,T).
The magnetic entropy approaches $\approx 99\%$ of the theoretically expected $R\ln8$ when increasing temperature to $T \approx 30$\,K.
The magnetic field dependence of the entropy in the AFM phase was found to be negligibly small when compared to  its temperature dependence (for moderate fields of $\mu_0H \leq 0.1$\,T).
\mat~is a promising ADR material that offers a high entropy density of 210\,mJ\,K$^{-1}$cm$^{-3}$ and minimum ADR temperatures lying well below its magnetic ordering temperature.
While it may not achieve the extremely low temperatures required for quantum computing with superconducting qubits (well below 100\,mK\,\cite{clarke_superconducting_2008}), it holds significant potential in typical two-stage or multi-stage ADR setups\,\cite{shirron_optimization_2014, prouve_athena_2020}. 
In such setups, one stage is designed to provide high cooling power and large heat capacity (precooling stage), while the other is optimized for reaching the lowest temperatures. 
\mat~appears particularly well-suited for the precooling stage.

\section*{Acknowledgments}
We thank Alexander Herrnberger and Klaus Wiedenmann for technical support and Unnikrishnan Arjun for collaborative work on NaYbP$_2$O$_7$. 
This work was supported by the Deutsche Forschungsgemeinschaft (DFG, German Research Foundation) - 514162746 (GE 1640/11-1) and TRR 360-492547816. 
PT acknowledges a postdoctoral fellowship from the Alexander-von Humboldt foundation.

We note that a German patent for the usage of $AB$P$_2$O$_7$ ($A$=alkaline metal, $B$=rare earth) for UHV compatible ADR to very low temperatures has been granted to the University of Augsburg (reference DE 10 2023 106 074 A1, March 10, 2023).
We further note that an international patent application under the PCT has been filed by the University of Augsburg for this patent with the WIPO (file reference PCT/EP2024/056108, March 7, 2024).
Inventors are Philipp Gegenwart, Unnikrishnan Arjun, and Anton Jesche. The remaining authors declare no competing interests.


%\bibliography{C:/Users/jeschean/Documents/Presentations/Publications/zitate}

\end{document}